\documentclass[amsmath,amssymb,aps,prb,reprint,floatfix,superscriptaddress]{revtex4-2}

\usepackage{graphicx}
\usepackage{dcolumn}
\usepackage{bm}
\usepackage{braket}

\usepackage{float}
\usepackage{hyperref}

\begin{document}

\title{Mott transition and correlation effects on strictly localized states in an octagonal quasicrystal}

\author{Efe Yelesti}
\email{efe.yelesti@bilkent.edu.tr}
\affiliation{Department of Physics, Bilkent University, Ankara, 06800, T\"urkiye}

\author{Onur Erten}
\affiliation{Department of Physics, Arizona State University, Tempe, AZ 85287, USA}

\author{M. \"O. Oktel}
\affiliation{Department of Physics, Bilkent University, Ankara, 06800, T\"urkiye}

\begin{abstract}
Flat-band systems have attracted significant attention as platforms for studying strongly correlated electron physics, where the dominance of electron-electron interactions over kinetic energy gives rise to a variety of emergent phenomena. Quasicrystals are compelling systems for studying these phenomena as they host degenerate strictly localized states at zero energy due to perfect destructive interference patterns. In this study, we use the slave-rotor mean-field approach to investigate the effects of electron interactions within the Hubbard model on the Ammann-Beenker quasicrystal. The phase diagram characterizing metallic and Mott insulator regions indicates a first-order phase transition. Our analysis shows that the local coordination number affects the local quasiparticle weight, displaying varying metallicity across the sites. Furthermore, we focus on the strictly localized states that arise in the non-interacting limit. We find that interactions and deviation from particle-hole symmetry induce spectral splitting, broadening, and partial delocalization of the localized states, depending on the local environment. In particular, 
certain localized states with higher coordination numbers remain more robust compared to others. Our results highlight the critical role of local geometry in shaping correlation effects in flat-band quasicrystals.

\end{abstract}

\maketitle
\section{Introduction}
In flat-band systems, the kinetic energy is highly suppressed which allows Coulomb interactions to dominate. This can give rise to a variety of strongly correlated phases including unconventional superconductivity, anomalous quantum Hall effect and Wigner crystals, to name a few\cite{Steglich_PRL1979,MacDonald_Rev2023,Haldane_PRL1983,Halperin_PRL1984,Wigner_PR1934,Wang_Nature2021}. Most studies have focused on translationally invariant systems such as kagome and Lieb lattices\cite{Lieb_PRL1989,Lee_Science2015,Mielke_1991,Wilson_PRM2021} and more recently moir\'e superlattices\cite{moire,Fernandes_PRB2018,Dean_Nature2019,Vishwanath_PRX2018,Efetov_Nature2019}, where momentum-space band structure provides a natural framework for understanding correlations. Recently, disordered and amorphous systems have also received attention for their unconventional localization and interaction effects \cite{Efros_Chapter1985,Atkinson_PRB2008,Altshuler_2006,Sarachik_2003}. Positioned between these extremes, quasicrystals offer a unique playground: they host long-range order without periodicity, allowing for a rich interplay between geometry and interaction-driven phenomena, and have increasingly attracted interest for exploring the role of Coulomb correlations in aperiodic environments \cite{Koga_2021,Koga_Penrose2015,Koga_Penrose2017,Andrade_PRB2019,Arita_PRR2019,Duncan_PRA2019,Derrico_PRL2024,Duneau_PRB2007,Milat_PRB2005,Tezuka_PRA2010,Koga_PRB2020,Hiramoto_1990,Fernandes_RX2024}.

\begin{figure}[t]
\includegraphics[scale = 0.27]{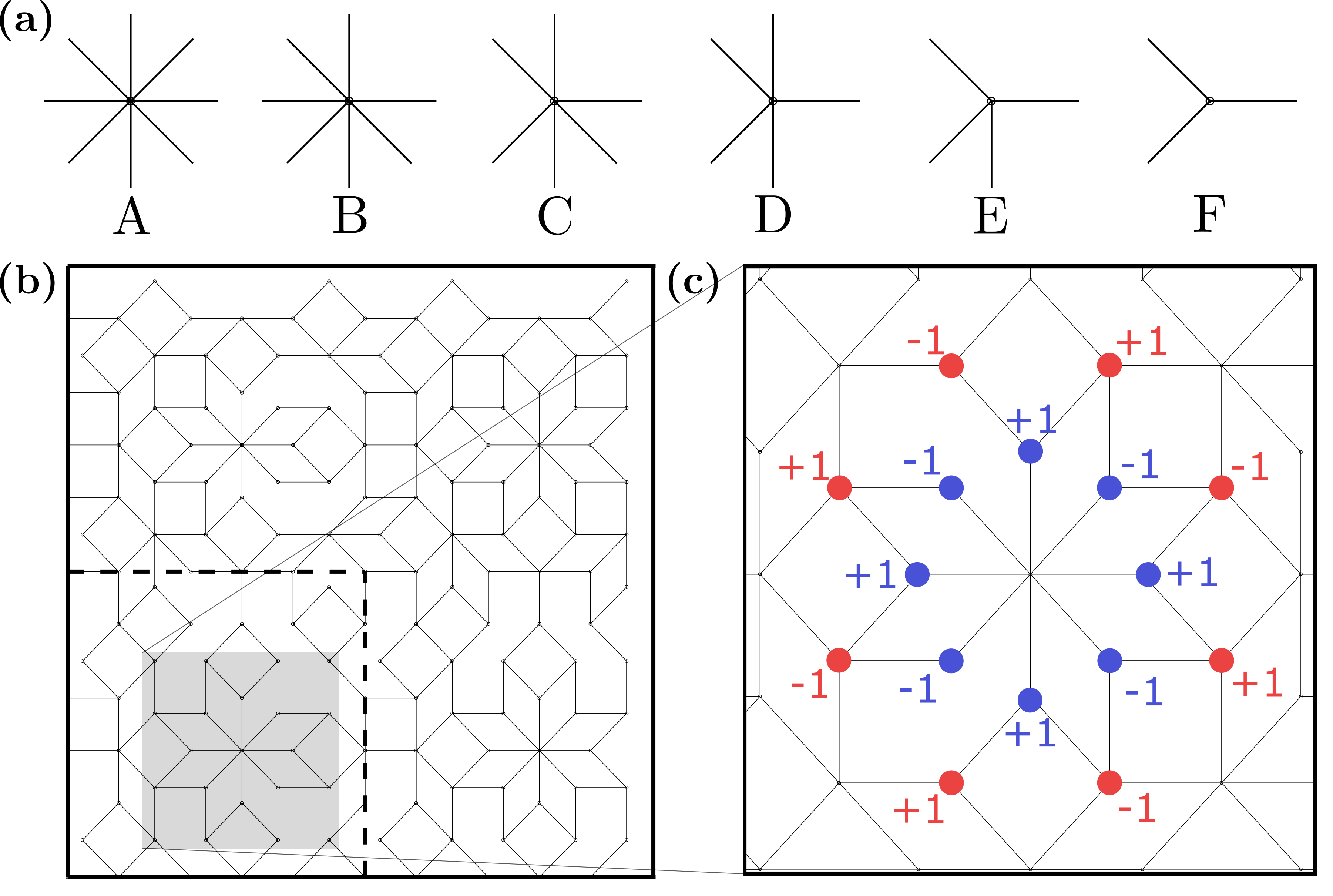}
\caption{(a) Vertex types A-F of the AB tiling with coordination numbers decreasing from 8 to 3. (b) Bipartite unit cell constructed from repeated four unit cells of the AB tiling approximant with $n = 2$ ($N_2 = 164$). (c) Strictly localized states of Type-A (blue) and Type-B (red) positioned in the nearest and next nearest neighbor of an A vertex in the AB tiling, where the electron wavefunction takes values of $\pm1$ on specific sites and undergoes destructive interference outside these sites. \label{Fig:"ABL1"}
}
\end{figure}

Unlike the infinitely repeating patterns of periodic structures, quasicrystals exhibit both translational and rotational order without their patterns repeating themselves at regular intervals, a feature that also sets them apart from glassy systems \cite{Levine_PRL1984}. The five- and eight-fold rotational symmetries, forbidden in periodic crystals but observed in early quasicrystals \cite{Shechtman_PRL1984}, are geometrically encoded in mathematical tilings such as the Penrose \cite{Bruijn_1981, Penrose_1974} and Ammann–Beenker (AB) patterns  \cite{Beenker_1982}. These distinct aperiodic configurations of lattice sites host strictly localized states (LS) at the zero-energy, formed through destructive interference at the boundary of their domain \cite{Koga_PRB2020,Oktel_PRB2021}. The macroscopic number of zero-energy states observed in these systems gives rise to a divergent peak at zero energy in the density of states, making them a key point for the study the effects of strong correlations.

Quasicrystals, first discovered in intermetallic alloys \cite{Shechtman_PRL1984}, displaying long-range order without periodicity, and have been realized across various physical platforms, including photonic lattices \cite{Vardeny_NaturePhotonics2013, Chan_PRL1998}, ultracold atoms \cite{Duneau_2014,Santos_PRA2005,Weld_PRA2015,Schneider_PRL2019} and moire systems \cite{Yao_Proceedings2018, Uri_Nature2023}. These systems have enabled direct probing of quasicrystalline symmetry, spectral properties, and localization phenomena.  Recently, magnetic order have been observed in alloy quasicrystals \cite{Deguchi_NatureMat2012,Tamura_ACS2021,Tamura_NaturePhysics2025}, and superconductivity has been reported in both 2D layered moiré systems and alloy quasicrystals \cite{Kamiya_NatComm2018,Uri_Nature2023,Tokumoto_NatComm2024}.

Signatures of electron correlation effects, such as anomalous transport and magnetism, have been theoretically studied in quasicrystalline structures \cite{Keskiner_PRB2025, Koga_2021,Koga_Penrose2015, Koga_Penrose2017, Andrade_PRB2019, Arita_PRR2019, Duncan_PRA2019, Duneau_PRB2007, Milat_PRB2005, Tezuka_PRA2010, Koga_PRB2020, Hiramoto_1990}, raising the question of how strong interactions manifest in aperiodic geometries. 

In this work, we study the Mott transition and effects of correlations on the strictly localized states in the AB tiling using the slave-rotor mean-field approach \cite{Georges_PRB2004}. Our main results are the following: (i) We obtain the phase diagram as a function of interactions and chemical potential which shows a first order metal to Mott insulator phase transition. (ii) We demonstrate the spatial distribution of the local quasiparticle weight and self energy across AB tiling in line with the local coordination number. (iii) We show the evolution the localized states with correlations. We report energy splitting of the type-A and type-B LS. Further, we investigate the hybridization with bulk states leading to partial delocalization of LS. 

The rest of the article is organized as follows. We begin by introducing the geometry and symmetry properties of the AB tiling, along with the construction of rational approximants via the cut-and-project method \cite{Duneau_1989, Duneau_RX2024}. We then define a Hubbard model on AB tiling and describe the slave-rotor approximation to study the effects of interactions. By solving the resulting self-consistent equations, we map out the phase diagram and analyze spatial variations of local observables such as the quasiparticle weight. Next, we characterize the energy shifts, broadening, and spatial delocalization of some of the strictly localized states, highlighting how local geometry modulates their stability against interactions. We conclude with a summary of our results and an outlook.

\section{Ammann-Beenker Tiling and Rational Approximants}
\label{sec:ABL_section}

Quasicrystals exhibit long-range order without translational invariance, positioning them between periodic crystals and disordered systems. Among various quasiperiodic tilings, the AB tiling stands out due to its geometric simplicity, eightfold rotational symmetry, and rich spectral properties \cite{Beenker_1982}. The AB tiling is composed of two tiles—a square and a rhombus with an internal angle of $\pi/4$—arranged such that all edges align along four distinct directions. These directions, called star vectors, are given by
\begin{equation}
    \hat{e}_0 = \hat{x}, \quad
    \hat{e}_1 = \frac{1}{\sqrt{2}}(\hat{x} + \hat{y}), \quad
    \hat{e}_2 = \hat{y}, \quad
    \hat{e}_3 = \frac{1}{\sqrt{2}}(-\hat{x} + \hat{y}),
\end{equation}
and every vertex in the AB tiling can be written as
\begin{equation}
    \vec{R}_{\text{AB}} = \sum_{j=0}^3 k_j \hat{e}_j, \quad \text{with } k_j \in \mathbb{Z}.
\end{equation}
This full set densely fills the plane, but only a subset corresponds to actual vertices of the quasicrystal, selected via a geometric constraint.

The AB tiling is generated through the cut-and-project method by selecting points from a four-dimensional hypercubic lattice $\mathbb{Z}^4$ that fall within a specific two-dimensional slice \cite{Beenker_1982}. A pair of mutually orthogonal planes is chosen in $\mathbb{R}^4$: a physical space $E$, which hosts the projected quasilattice, and a complementary perpendicular space $E'$ used for selection. The orientation of these planes is fixed by the eigenspaces of a rotation matrix $\bm{g} \in \mathbb{R}^{4\times 4}$ with eigenvalues $\{e^{i\pi/4}, e^{i3\pi/4}, \text{c.c.}\}$. Projection matrices $\bm{\pi}$ and $\bm{\pi'}$ map lattice points from $\mathbb{Z}^4$ into $E$ and $E'$, respectively. The real-space structure of the AB tiling is determined by projecting those lattice points whose perpendicular components lie inside a compact window $W \subset E'$, which takes the shape of a regular octagon \cite{Beenker_1982}.

This construction produces a deterministic, non-periodic tiling with local environments varying in coordination number from 3 to 8. The lattice is bipartite: its sites can be divided into two sublattices such that all nearest-neighbor connections occur between opposite sublattices. This structure gives rise to a particle-hole symmetric spectrum in tight-binding models. One notable spectral feature is a highly degenerate set of strictly localized states (LS) at zero energy. These LS are confined to a finite number of sites, on a single sublattice, and result from destructive interference enforced by the underlying quasiperiodic geometry. We will focus particularly on the most common types of such states, referred to as type-A and type-B, which are centered around vertices with high local symmetry.

For numerical calculations, we work with periodic rational approximants to the AB tiling, which offer a way to study large finite systems while minimizing edge effects \cite{Duneau_1989, Duneau_RX2024}. These are constructed by slightly modifying the orientation of the projection planes so that the irrational components become rational approximations. Specifically, the silver mean $\lambda = 1 + \sqrt{2}$ is replaced by successive Pell ratios $\mathcal{P}_n / \mathcal{P}_{n+1}$, with $\mathcal{P}_n$ denoting the $n$th Pell number. The modified projection yields a periodic tiling with square unit cells, whose size grows rapidly with $n$. To preserve the tile shapes and symmetry as closely as possible, a linear transformation is applied before projection \cite{Duneau_1989, Duneau_RX2024}. The resulting unit cells retain approximate eightfold symmetry and provide an efficient and accurate way to sample the bulk behavior of the infinite quasicrystal. The unit cells constructed using this method, which can be repeated periodically, are shown in Fig.~\ref{Fig:"ABL1"}. 

\section{Model and methods}
We consider a single-band Hubbard model defined on AB tiling verticies, $\mathcal{H}=\mathcal{H}_0+\mathcal{H}_U$,
\begin{eqnarray}
    \mathcal{H}_0 &=&-t\sum_{\langle i,j \rangle,\sigma} d^{\dagger}_{i\sigma} d_{j\sigma} -\mu \sum_{i\sigma} d^{\dagger}_{i\sigma} d_{i\sigma}  \\
    \mathcal{H}_U& =& \frac{U}{2}\sum_i\left[\sum_\sigma d_{i\sigma}^{\dagger} d_{i\sigma}-1\right]^2 \label{eq:"H_U"}
\end{eqnarray}
where $d^{\dagger}_i$ is fermionic creation operator, $\mu$, $t$ and $U$ are the chemical potential and nearest neighbor hopping amplitude and onsite Coulomb repulsion respectively, $\sigma = 1,\cdots,N$ is the spin/orbital index for which we consider $N=2$ for single band Hubbard model.  

\begin{figure}[!t]
\includegraphics[scale = 0.3]{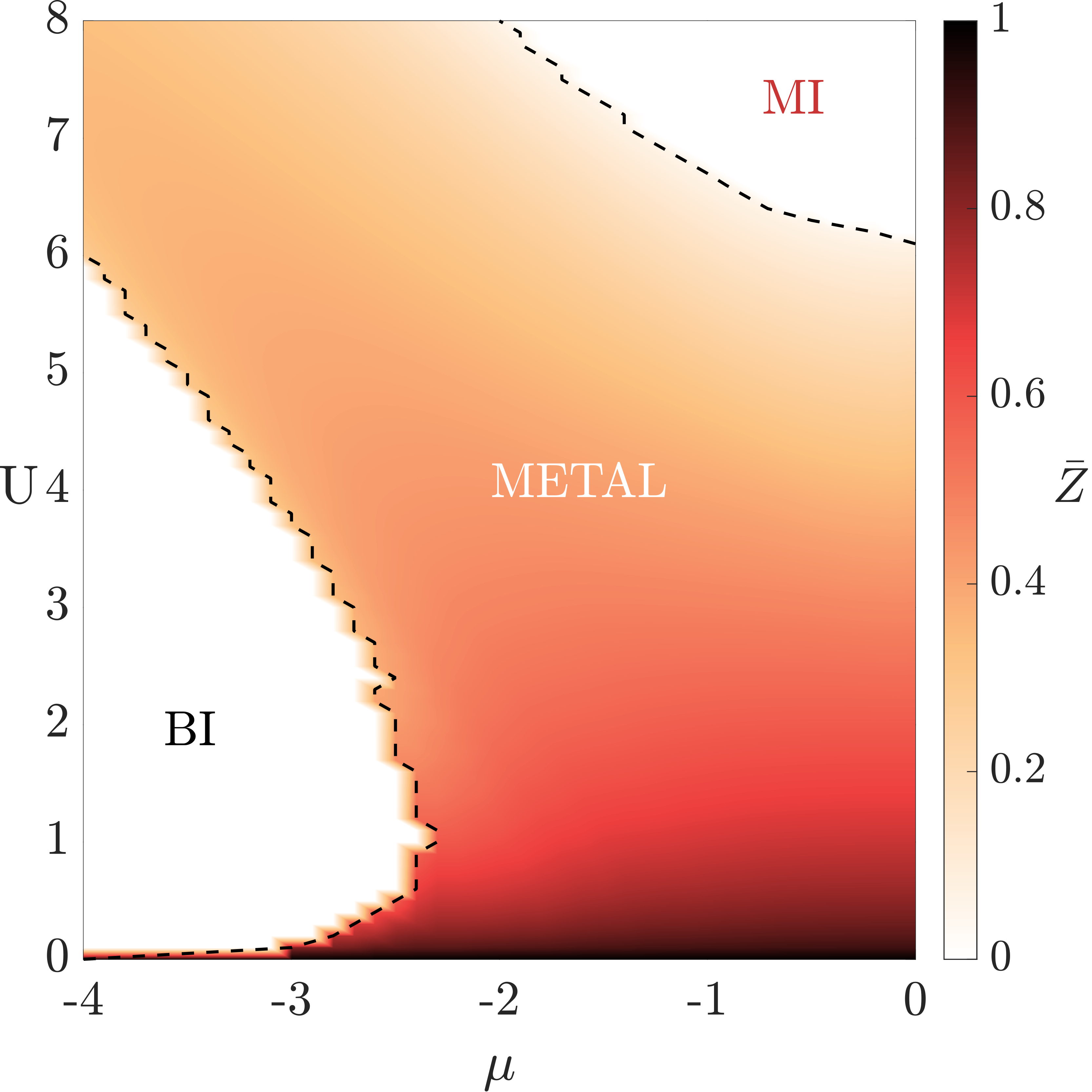}
\caption{
Slave-rotor mean field phase diagram of single band Hubbard model on AB tiling as a function of U and $\mu$. The system undergoes a metal-Mott insulator (MI) phase transition as the interaction strength increases at half-filling. As the chemical potential increases the system becomes band insulator (BI). The colorbar represents the average quasiparticle weight, $\bar{Z}$. \label{Fig:"Phase1"}}
\end{figure}

Next, we discuss the slave-rotor mean-field formalism\cite{Georges_PRB2002, Georges_PRB2004}. Originally introduced as an impurity solver\cite{Georges_PRB2002}, slave-rotor approach is able to capture essential aspects of the non-magnetic metal to Mott insulator transition including frequency dependent, imaginary self-energy and Hubbard subbands. In addition, it is easier to include spatial fluctuations and perform larger scale calculations compared to more numerically expensive methods such as quantum Monte Carlo and dynamical mean-field theory. It has been applied to frustrated Hubbard models\cite{Zhao_PRB2007}, correlated superconductors\cite{Ho_PRB2011} and multiband systems\cite{Meetei_PRL2013, Banerjee_NatPhys2013} to study the effects of correlations.

We start by introducing a neutral spinon that carries the spin degrees of freedom (DOF) and a bosonic O(2) rotor that is associated with the total charge. Hence the original fermionic operator is fractionalized as $d_{i\sigma}^\dagger = f_{i\sigma}^\dagger e^{i\theta_i}$. The local fermionic Hilbert space, originally spanned by the fermionic operators $d_{i\sigma}^{\dagger}$, is enlarged as $\mathcal{H}_d^i \rightarrow \mathcal{H}_{\theta}^i \otimes \mathcal{H}_{f}^i$. To project back onto the physical Hilbert space, we impose a local constraint that links the spinon number to the rotor's angular momentum
\begin{equation}\label{eq:constraint}
    L_i = \sum_{\sigma} \left(f^\dagger_{i,\sigma}f_{i,\sigma} - \frac{1}{2}\right)
\end{equation}

\begin{figure}[!t]
\includegraphics[scale = 0.2752]{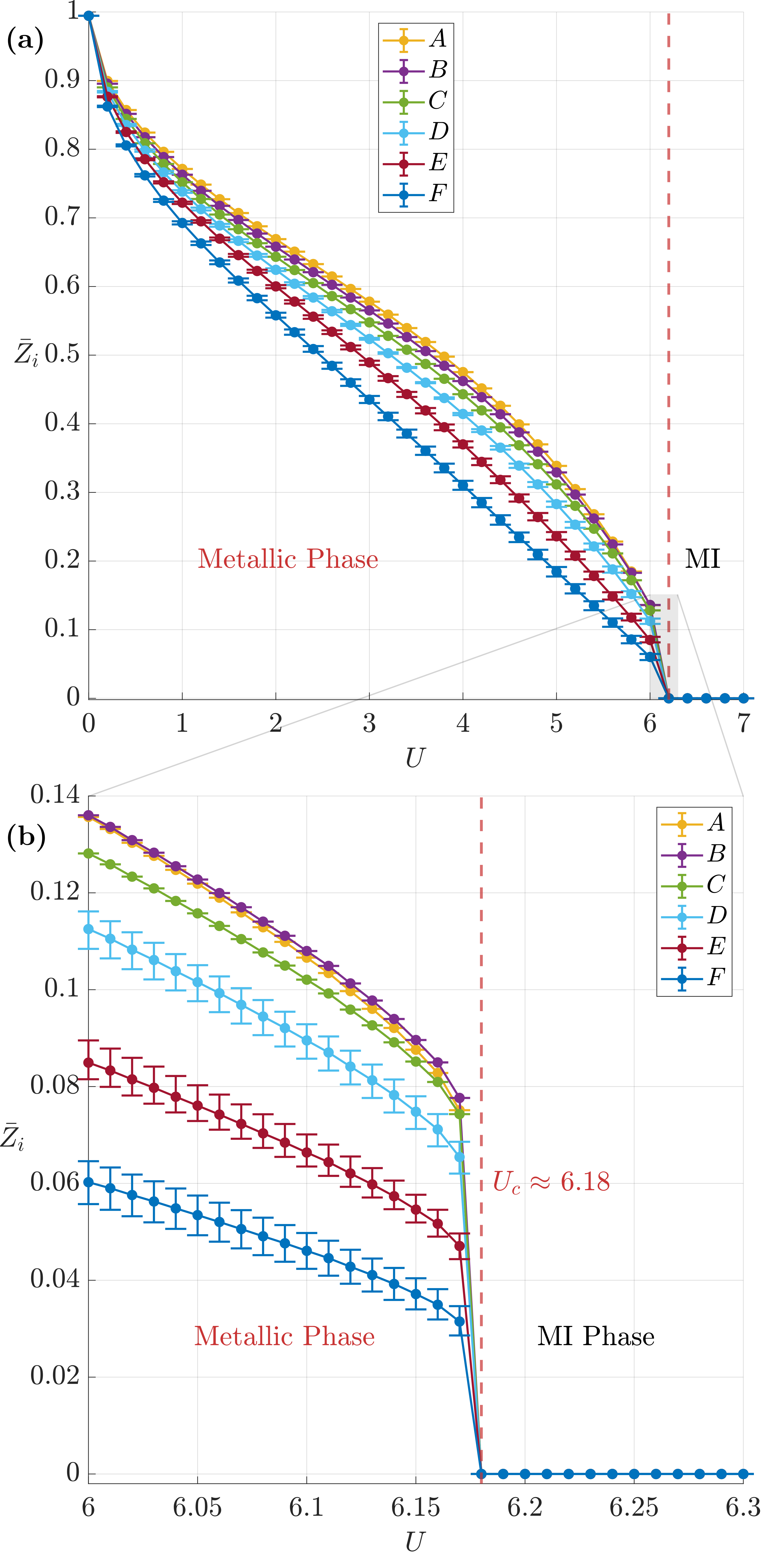}
\caption{Average quasiparticle weight, $\bar{Z}$ as a function of U for different types of local coordination that are depicted in Fig.~\ref{Fig:"ABL1"} for $\mu=0$. The jump $\bar{Z}$ signifies a first order metal-MI transition between $U = 6.17$. Different colored lines represent the averages of $\bar{Z}_i$ over lattice sites that share the same coordination number. Average local quasiparticle weight $\bar{Z}_i$ reflects the variations in local coordination number. Error bars indicate the range of $Z_i$ values for sites sharing the same coordination number. \label{Fig:"Phase2"}}
\end{figure}

The total angular momentum at each site $L_i$ must match the deviation of the fermion number from half-filling, ensuring that the total charge stored in the rotor matches the fermionic occupation at each site. Hence, the angular momentum is constrained to $\ell = (Q - 1)$, where $Q$ is the total charge. With the introduction of the bosonic rotor variable \(\theta\), with conjugate angular momentum operator \(\hat{L} = -i \partial/\partial \theta\), the original quartic Hubbard interaction transforms into a quadratic term, $\mathcal{H}_U = \sum_i \frac{U}{2} L_i^2$. To ensure that the slave-rotor decomposition faithfully represents the original physical system, the constraint must be imposed directly on the Hamiltonian. This is done by introducing a Lagrange multiplier $h_i$ at each site, which dynamically enforces the relation between rotor angular momentum and spinon number. The constraint term takes the form

\begin{equation}
    \mathcal{H}_h = \sum_i h_i \left( \hat{L}_i - \sum_\sigma \left\{f_{i\sigma}^\dagger f_{i\sigma} - \frac{1}{2}\right\} \right)
\end{equation}
giving the full Hamiltonian $\mathcal{H} = \mathcal{H}_0+\mathcal{H}_U+\mathcal{H}_h$.

To proceed, we employ a mean-field decoupling of the slave-rotor Hamiltonian, which allows us to separate it into two coupled parts: one describing the dynamics of the spinons, and the other governing the rotor degrees of freedom. We write operators in terms of their averages and fluctuations, and retain only the leading-order terms. In addition, the average values are assumed to be real on the bonds (i.e. no time reversal symmetry breaking). This yields a simplified effective Hamiltonian consisting of two coupled mean-field parts: a spinon Hamiltonian with rotor-renormalized hopping, and a rotor Hamiltonian with spinon correlations entering as an external field. To further simplify the problem, we apply a second mean-field approximation to the rotor part. This leads to a simplified single-site Hamiltonian for the rotor at each site. The decoupled Hamiltonians are as

\begin{align}\label{eq:hrotor_spinon}
\mathcal{H}_{f} &= -\sum_{i,\sigma} (\mu+h_i)f^\dagger_{i,\sigma}f_{i,\sigma} - \sum_{i,j,\sigma} t_{ij}^{\text{eff}}f^\dagger_{i,\sigma}f_{j,\sigma} \nonumber \\
\mathcal{H}_\theta &= \sum_i \left( \frac{U}{2} \hat{L}_i^2 + h \hat{L}_i + K_i \cos \theta_i \right)
\end{align}
where the effective parameters $t_{ij}^{\text{eff}}$ and $K_i$ are given as 
\begin{align} \label{eq:"CoupledParameters"}
    K_i &= -2 \sum_j \left( \sum_\sigma t_{ij} 
    \left\langle f^\dagger_{i\sigma} f_{j\sigma} \right\rangle_f \right) 
    \left\langle \cos \theta_j \right\rangle_\theta \nonumber \\
    t_{ij}^{\text{eff}} &= t_{ij} 
    \left\langle \cos\theta_i \right\rangle_\theta \left\langle \cos\theta_j \right\rangle_\theta 
\end{align}

Here, we define the local quasiparticle weight by $Z_i = \langle \cos(\theta_i) \rangle^2$. When $Z_i$ is finite, the system is metallic, and the effective mass of the electron is renormalized as $m^*/m \sim 1/Z_i$. As interactions increase, $Z_i$ decreases and the electron becomes heavier. At the Mott transition, $Z_i \to 0$, the effective mass diverges, and the electron becomes completely localized. Thus, tracking $Z_i$ provides a clear diagnostic of the Mott transition. 

\begin{figure}[!t]
\includegraphics[scale = 0.22]{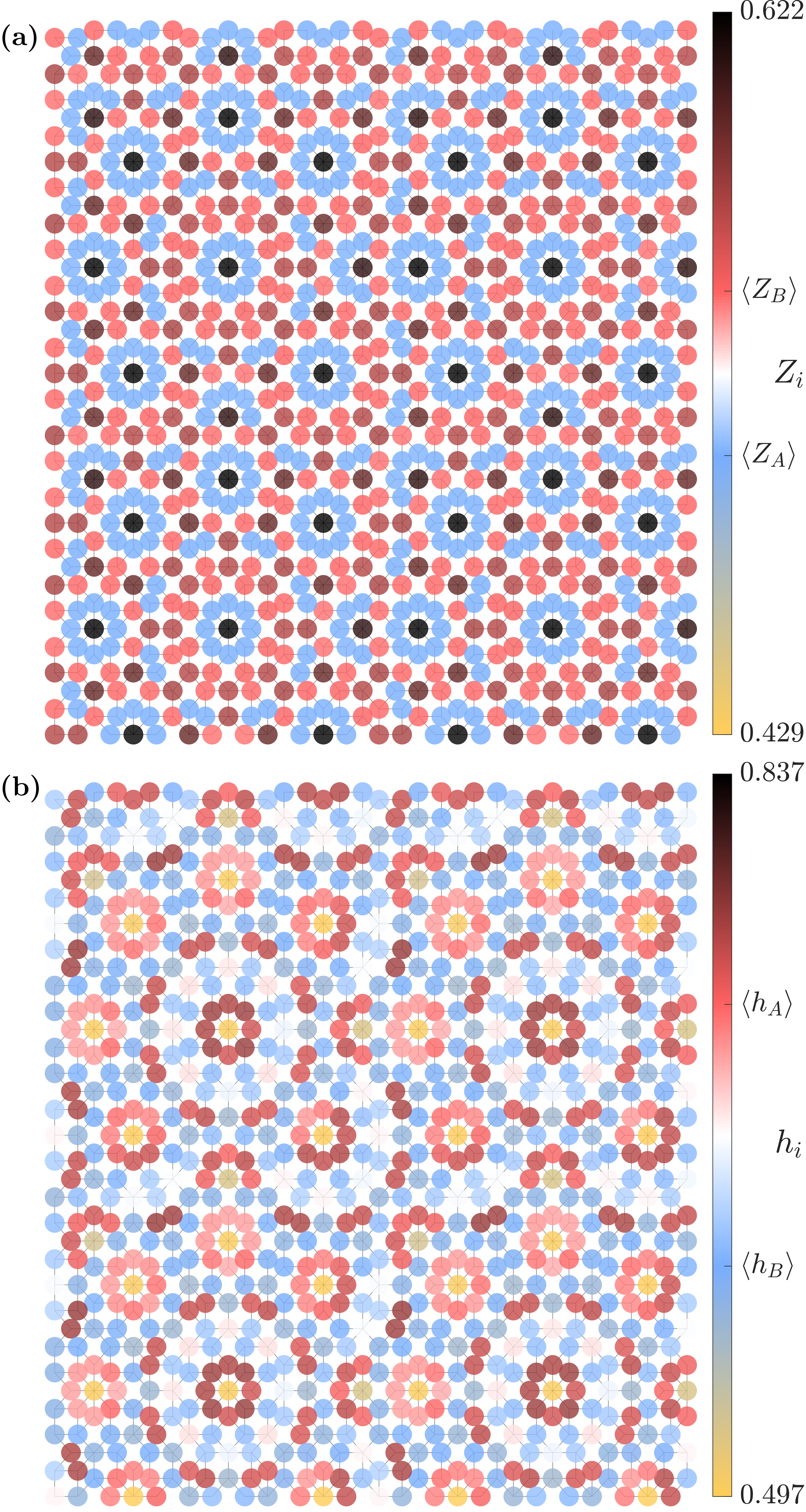}
\caption{Spatial distribution of local observables on the AB tiling for \(\mu= -1\) and \(U = 2.5\). (a) Local quasiparticle weight \(Z_i\) and (b) Lagrange multiplier field \(h_i\) both show spatial inhomogeneity due to the aperiodic geometry. $\langle h_{A(B)} \rangle$ and $\langle Z_{A(B)} \rangle$ represent the average value of $h$ and $Z$ for A(B)-type LS.
\label{Fig:"h_Z_local"}}
\end{figure}

Using this self-consistent mean-field framework, we compute the local quasiparticle weights $Z_i$ across a range of interaction U and $\mu$. Averaging over all sites, we define $\bar{Z} = \frac{1}{N} \sum_i Z_i$ as an indicator of global metallicity. A finite $\bar{Z}$ signals the presence of coherent quasiparticles and a metallic phase, while $\bar{Z} \to 0$ marks the onset of the Mott localization. 
\section{Results and discussion}
\subsection{Mott transition}

We present the self-consistent mean field phase diagram as a function of $U$ and $\mu$ in Fig.~\ref{Fig:"Phase1"} which reveals three distinct regions: a correlated metal phase at intermediate interactions, a Mott-insulating (MI) phase at large $U$ at half-filling, and a band insulating (BI) phase at for large $|\mu|$. The metal-MI transition is driven by increasing interaction strength, whereas BI phase occurs when the spinon bands become fully occupied or empty. 

\begin{figure}[!t]
\includegraphics[scale = 0.22]{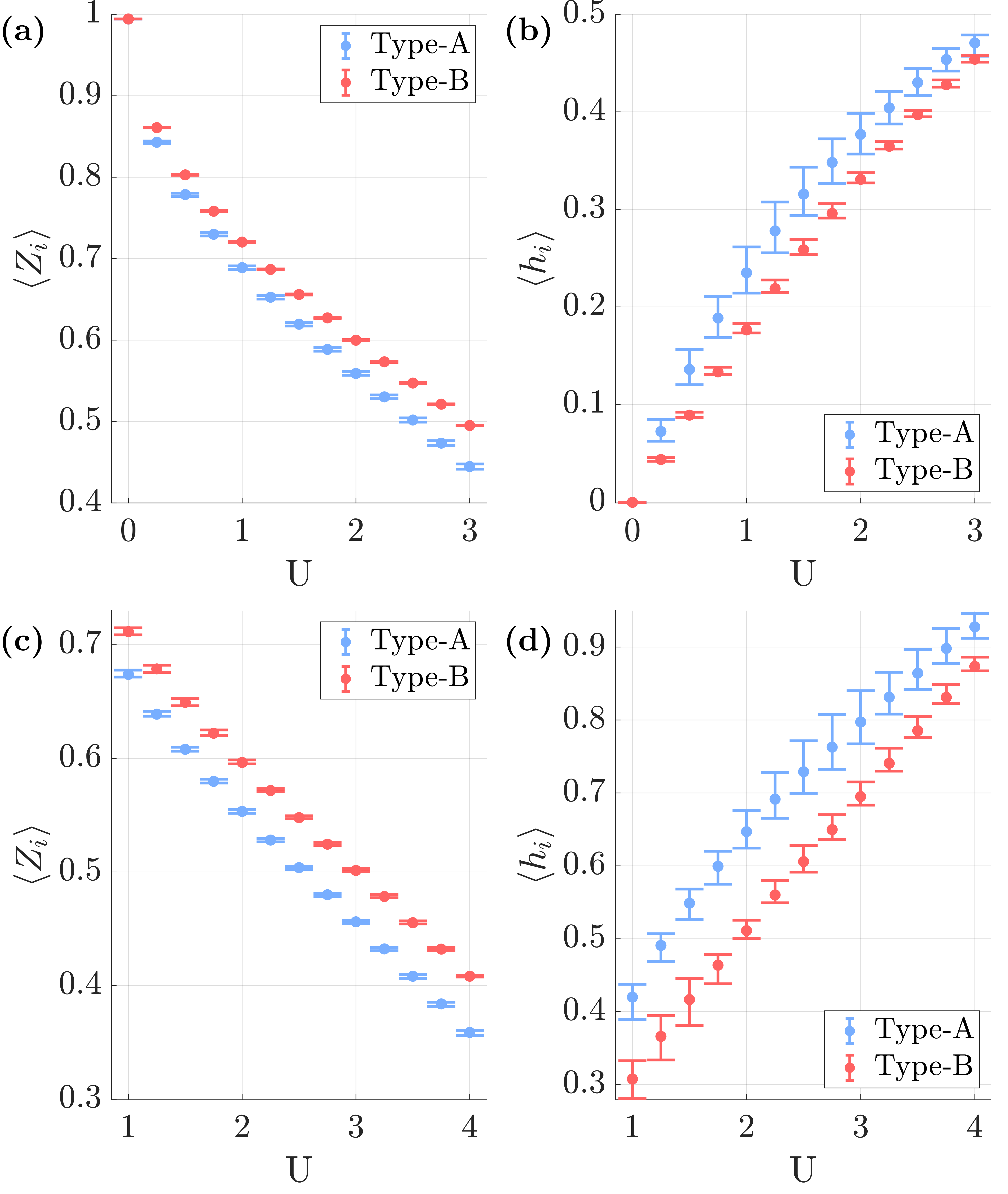}
\caption{
Interaction dependence of local observables for Type-A and Type-B localized states at two different values of the chemical potential: (a,b) $\mu = -0.5$, (c,d) $\mu = -1.0$. (a,c) shows the local quasiparticle weight $Z_i$, and (b,d) show the corresponding Lagrange multiplier $h_i$, both presented as functions of interaction strength $U$. Blue and red data points correspond to averages over sites identified with Type-A and Type-B localized states, respectively. Error bars indicate the range of values within each type, capturing the inhomogeneity due to the underlying lattice structure. \label{Fig:"h_Z_all"}
}
\end{figure}

To further examine the nature of the Mott transition in the AB tiling, we analyze the spatial distribution of the local quasiparticle weights $Z_i$ as a function of interaction strength $U$ at particle-hole symmetry ($\mu = 0$). As shown in Fig.~\ref{Fig:"Phase2"}, the values of $Z_i$ vary significantly across sites within the metallic phase, reflecting the inherent spatial inhomogeneity induced by the aperiodic lattice structure. Sites with higher coordination tend to exhibit larger $Z_i$. Similar behavior has been observed in dynamical mean field theory \cite{Koga_Penrose2015} and quantum Monte Carlo\cite{Hiramoto_1990} studies of Mott transition in other quasicrystals such as Penrose tiling. As the interaction approaches the critical value \(U_c \approx 6.17\), all \(Z_i\) values collapse sharply to zero, consistent with a first-order phase transition. In Fig.~\ref{Fig:"h_Z_local"}(a), we present the real space distribution of $Z_i$ that provides a visual demonstration of our results in Fig.~\ref{Fig:"Phase2"} for a given $\mu$ and $U$. The presence of coordination-dependent \(Z_i\) distributions underscores the role of local geometry in modulating electronic coherence in quasicrystalline systems. While the Lagrange multiplier, $h_i$, is strictly equal to zero for $\mu=0$ due to particle hole symmetry, it is non-zero and site dependent for finite $\mu$, similar to $Z_i$ as shown in Fig.~\ref{Fig:"h_Z_local"}(b). Further, regions of reduced \(Z_i\), which indicate suppressed quasiparticle coherence, tend to correlate with higher \(h_i\) values, consistent with the interpretation of \(h_i\) as an effective potential penalizing charge fluctuations. Furthermore, this modulation has a correlation with the structural classification of localized states: sites associated with Type-A LS typically exhibit lower \(Z_i\) and higher \(h_i\), whereas Type-B LS centers remain relatively more metallic.

Having established the Mott transition and the overall phase structure of the interacting system, we now turn our attention to the behavior of LS in the metallic phase of the AB tiling. These states are strictly localized in the non-interacting limit, arising from the quasiperiodic geometry of the lattice. However, it remains an open question how they are modified by electron correlations within the metallic regime, before the onset of full Mott localization. In the following section, we investigate how interactions affect the weight, spatial profile, and robustness of these states when quasiparticle coherence is still preserved.

\begin{figure*}
\includegraphics[scale = 0.23]{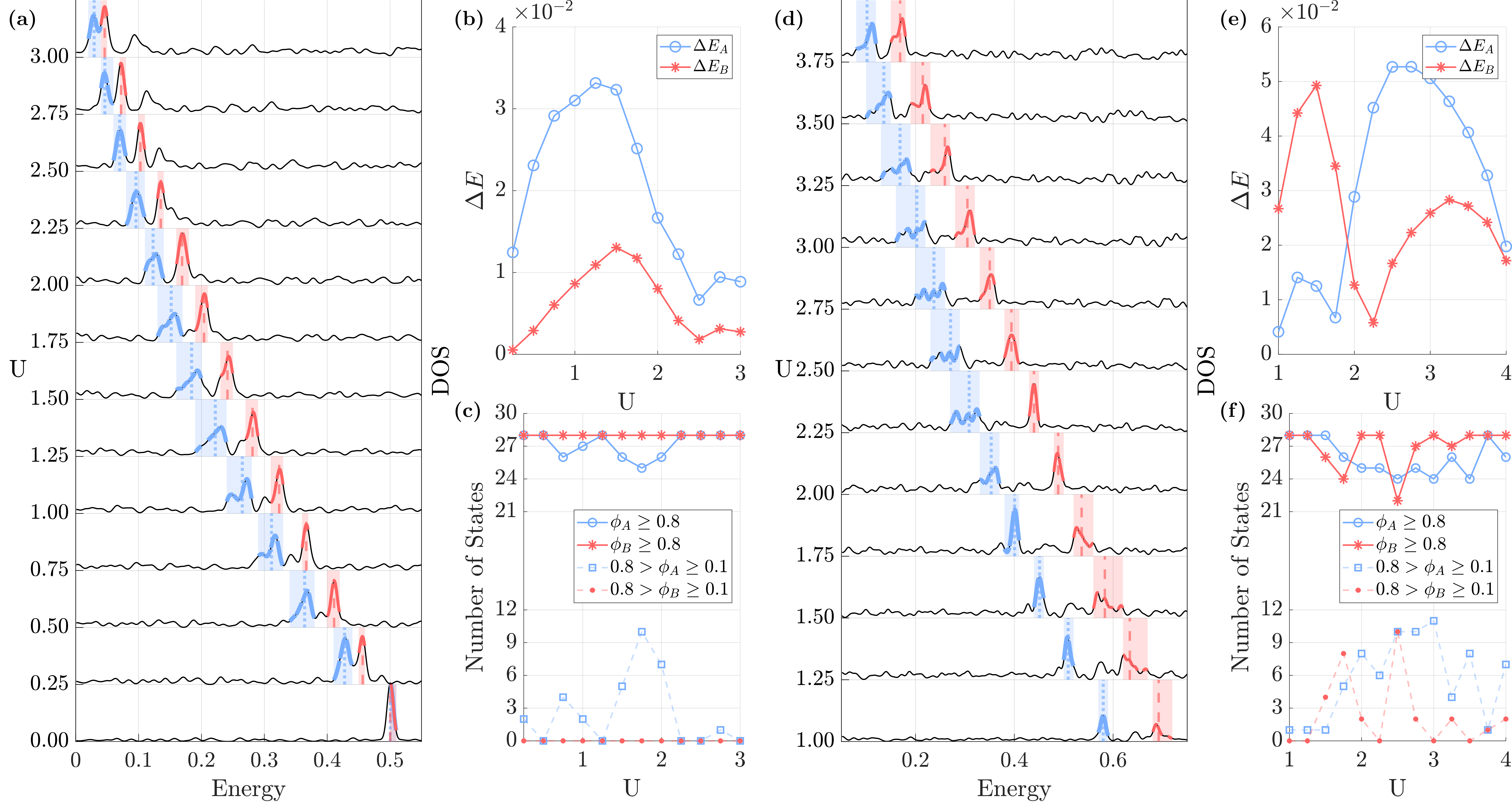}
\caption{Evolution of DOS for different chemical potentials $\mu = -0.5$ (a) and  $\mu = -1.0$ (d). States with high overlap with the Type-A and Type-B LS manifolds are highlighted in the plots, using blue and red, respectively. Each red and blue vertical lines indicate the expected energy of the corresponding localized state type, given by $E = -(\mu + \langle h_i\rangle)$ where $i$ runs over types A and B.  (a) For $\mu = -0.5$ and $U=0$, the DOS shows a peak at $E = 0.5$ originating from the presence of both types of localized states. (b) and (e) show the interaction-dependent broadening of the DOS peaks associated with Type-A and Type-B localized states, denoted by $\Delta E_A$ and $\Delta E_B$, respectively. The peak widths are extracted from the energy range of spinon eigenstates that have an overlap $\phi > 0.1$ with the non-interacting LS manifold of the corresponding type (A or B). These contributing states are also highlighted in the DOS plots (a, b) using bold blue and red vertical lines. (c) and (f) show the number of spinon states exhibiting different levels of overlap $\phi$ with the noninteracting LS manifolds. Here, $\phi_{A}$ and $\phi_{B}$ denotes the projection of a spinon eigenstate onto the Type-A (Type-B) LS basis. Markers distinguish states with high overlap $(\phi \geq 0.8$) and intermediate overlap ($0.8 > \phi \geq 0.1$). \label{Fig:"DOS_all"}}
\end{figure*}

\subsection{Effects of correlations on the localized states}
To better understand how electron correlations affect distinct classes of localized states, we focus on two representative types, Type-A and Type-B, identified from the non-interacting limit (see Fig.~\ref{Fig:"Phase1"}(d)). Since the coherent part of the Green's function in slave-rotor mean field theory is 
\begin{equation}
    G_{\rm coh}(i,j, \omega) = \frac{\cos(\theta_i)\cos(\theta_j)}{\omega+i\eta - H_f(i,j)}\label{eq:GF}
\end{equation}
where $H_f$ is the spinon Hamiltonian given in eq.~\ref{eq:hrotor_spinon}. As such, we analyze the single-particle spinon spectrum by examining the distribution of local observables such as the Lagrange multiplier $h_i$ and the quasiparticle weight $Z_i$.
These quantities provide insight into how interactions influence the coherence and localization properties of different LS types within the metallic regime.

In Fig. \ref{Fig:"h_Z_all"}, we compare the interaction dependence of the local quasiparticle weight $Z_i$ and the corresponding Lagrange multiplier $h_i$ for lattice sites associated with Type-A and Type-B localized states, at two representative chemical potentials: $\mu = -0.5$ and $-1.0$. As mentioned in the previous section $h_i$ is strictly zero for $\mu=0$ and the particle-hole symmetry allows for $\mu \rightarrow -\mu$ mapping. For each case, the average value is shown as a function of $U$, and the error bars reflect the spread across all sites belonging to a given type. In all cases, we observe that increasing interaction strength reduces $Z_i$ and enhances $h_i$, consistent with the expected suppression of charge fluctuations. Notably, Type-B sites, having a coordination number of 4 compared to 3 for Type-A, consistently exhibit slightly larger $Z_i$ and lower $h_i$. Their higher connectivity provides more hopping channels for charge fluctuations, making them less sensitive to interaction effects and more robust to sustaining metallic behavior. As a result, they retain a stronger metallic character.

From the form of the spinon Hamiltonian in eq. \ref{eq:hrotor_spinon}, we expect that interactions effectively shift the energies of the localized spinon states. In the non-interacting limit, these states appear as sharp peaks in the spinon density of states (DOS), centered at the chemical potential. Within the slave-rotor formalism, the effect of interactions enters through the Lagrange multipliers $h_i$, which acts as a real part of the self energy for the physical electron Green's function (eq.~\ref{eq:GF}). Consequently, the energies of localized states shift from $-\mu$ to $- (\mu+\langle h_i \rangle)$, with the average taken over the subset of lattice sites associated with a given LS type (Type-A or Type-B). 

This predicted shift is marked in Fig. \ref{Fig:"DOS_all"} with vertical blue and red lines, corresponding to $-(\mu+\langle h_i \rangle_A)$ and $-(\mu+\langle h_i \rangle_B$), respectively. Furthermore, the spread in $h_i$ values within each manifold contributes to the broadening or narrowing of the corresponding DOS peaks: a wider distribution in $h_i$ results in a broader peak, while a more uniform $h_i$ profile yields a sharper spectral feature. These trends are clearly visible in Fig. \ref{Fig:"DOS_all"}, which shows the evolution of the spinon DOS at two representative chemical potentials, $\mu = -0.5, -1.0,$ [Fig. \ref{Fig:"DOS_all"} (a), (d), respectively], as the interaction strength $U$ is varied. For $\mu = -0.5$ [Fig. \ref{Fig:"DOS_all"} (a)], the system begins with a single sharp peak at $E = 0.5$, corresponding to the degenerate LS manifold in the non-interacting limit. As $U$ increases, the peak splits into two components, reflecting the interaction-induced energy splitting between Type-A and Type-B states. Simultaneously, both peaks broaden significantly, signaling that interactions cause a partial delocalization of the LS. At $\mu = -1.0$ [Fig. \ref{Fig:"DOS_all"} (d)], a similar splitting and broadening occurs, although the peaks become more asymmetric, reflecting the different responses of each LS type to interaction.

To quantify the stability of the LS against interactions, we introduce an overlap-based metric that measures how closely the single-particle spinon eigenstates in the interacting system resemble the strictly localized states of the non-interacting limit. These non-interacting LS, categorized as Type-A and Type-B based on their spatial structure (see Fig.~\ref{Fig:"ABL1"}), span two orthonormal manifolds. We construct projectors onto these manifolds: \(P_A = \sum_i \ket{\psi^A_i}\bra{\psi^A_i}\) and \(P_B = \sum_i \ket{\psi^B_i}\bra{\psi^B_i}\), where \(\{\ket{\psi^A_i}\}\) and \(\{\ket{\psi^B_i}\}\) are the orthonormal bases for the Type-A and Type-B localized states, respectively.

Each eigenstate \(\ket{\xi_m}\) of the self-consistent solutions is analyzed by projecting it onto these manifolds. We compute the scalar overlap
\[
\phi = \braket{\xi_m|P|\xi_m},
\]
where m is running over all of the eigenstates and \(P\) is either \(P_A\) or \(P_B\), depending on the LS type under consideration. This overlap \(\phi\) gives a quantitative measure of how much of the eigenstate's weight resides within the corresponding LS manifold.

A value of \(\phi \approx 1\) indicates that the interacting eigenstate remains almost entirely within the original LS space—i.e., its character is preserved under interaction. Conversely, a small value of \(\phi\) signals that the state has significantly hybridized with extended or other delocalized modes. In our analysis, we use a practical threshold at \(\phi > 0.8\) to classify states as ``intact'' meaning that the interactions do not substantially distort their localized character or disrupt the underlying destructive interference pattern responsible for their localization. When \(\phi < 0.8\), we observe the onset of delocalization, where interactions begin to deform the LS and allow spectral weight to leak outside of the original subspace. This metric thus allows us to systematically track how LS evolve with interaction strength and chemical potential, revealing whether they persist, broaden, or dissolve into the continuum of metallic states. Fig.~\ref{Fig:"DOS_all"} (c,f) show, for each chemical potential, how the number of states with high overlap (\(\phi \geq 0.8\)) varies as a function of interaction strength \(U\). These counts provide a quantitative estimate of how many LS remain intact as interactions grow. We expect 28 possible localized states for each type, based on a direct count over the n=3 approximant (956 sites) bipartite tiling. In all three cases, the total number of intact LS remains approximately constant at weak and strong coupling limits, but shows a noticeable dip in an intermediate interaction window. The most pronounced suppression is observed for \(\mu = -1.0\) around \(U \approx 2.5\), where the number of states with \(\phi \geq 0.8\) decreases significantly and is partially replaced by states with intermediate overlap values (\(0.8 > \phi > 0.1\)). This trend suggests that LS in this regime are strongly hybridized with the surrounding states, leading to a partial breakdown of their localized nature. Comparing across different chemical potentials, we find that LS are most susceptible to delocalization in the intermediate \(\mu = -1.0\) case, while they remain more robust at (\(\mu = -0.5\)). This is consistent with the DOS analysis, where broader peaks and reduced \(\phi\) values accompany the LS deformation. 

\begin{figure}[!t]
\includegraphics[scale = 0.23]{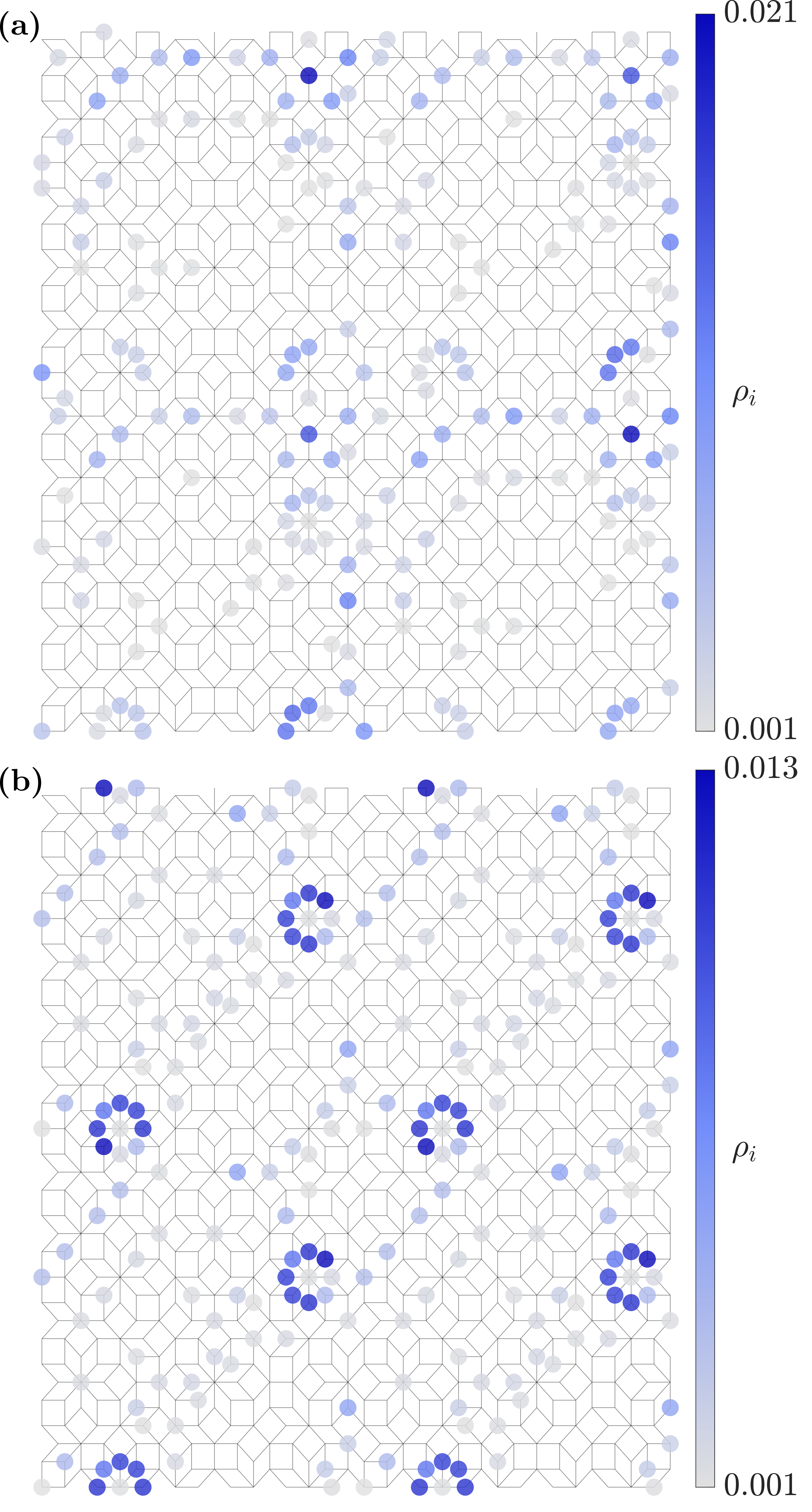}
\caption{\label{Fig:"LS_eg_A"} Real-space spinon density profiles \(\rho_i\) for four selected eigenstates at \(\mu = -1.0\), \(U = 2.5\), chosen to illustrate different degrees of overlap with the non-interacting Type-A LS manifold. (a) and (b) show states with low $\phi_A = 0.132$ and intermediate overlap $\phi_A = 0.513$.}
\end{figure}

\begin{figure}[!t]
\includegraphics[scale = 0.23]{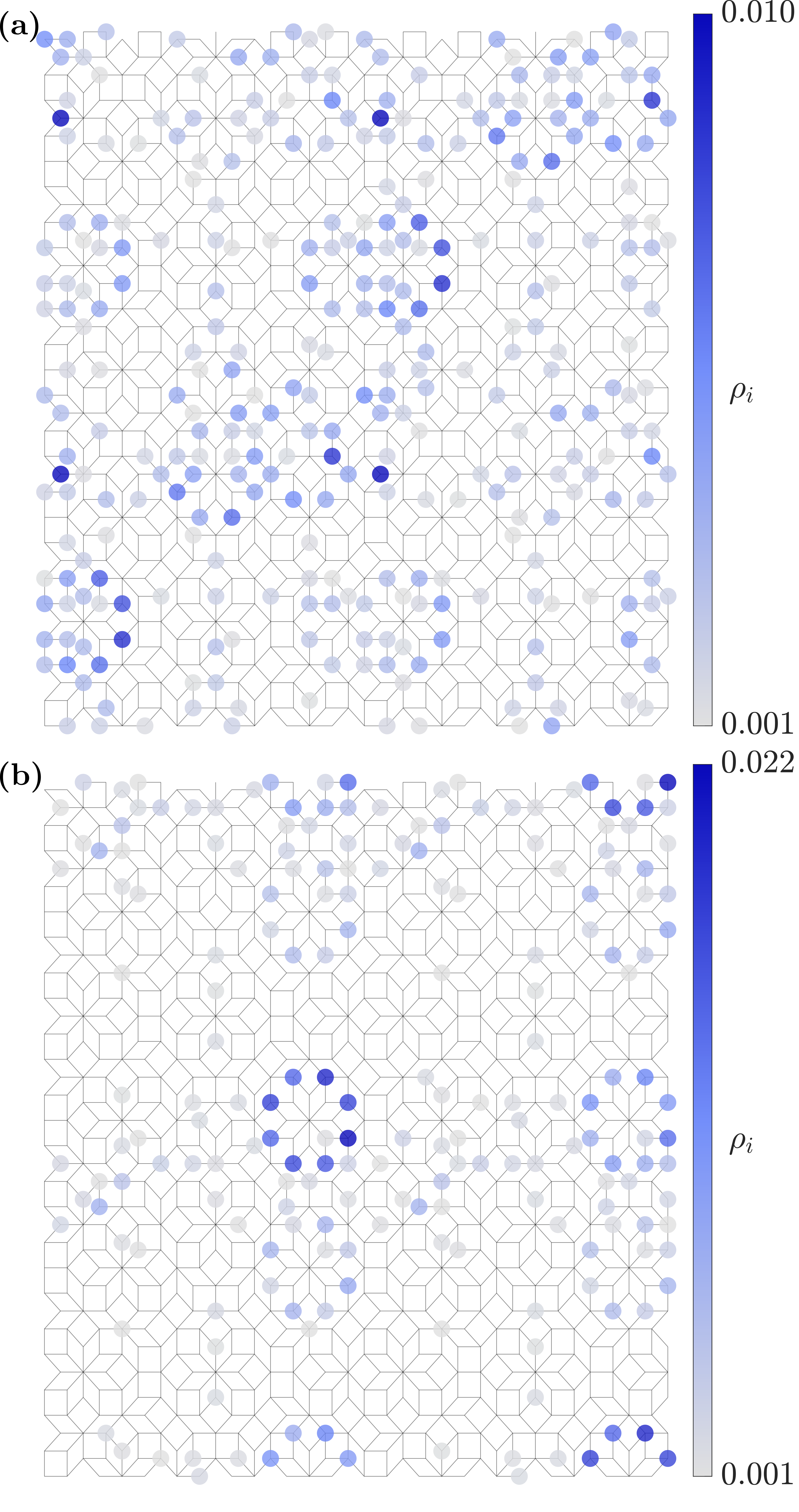}
\caption{\label{Fig:"LS_eg_B"} Real-space spinon density profiles \(\rho_i\) for two selected eigenstates at \(\mu = -1.0\), \(U = 2.5\), chosen to illustrate different degrees of overlap with the non-interacting Type-B LS manifold. (a) and (b) show states with low $\phi_B = 0.232$ and intermediate overlap $\phi_B = 0.575$.}
\end{figure}

In the non-interacting case, the AB tiling host strictly localized states with a compact support, i.e. wavefunction is exactly zero outside a bounded region. 

As $U \rightarrow 0$ the effective hopping strength that encodes the local structure information of the aperiodic structure $t_{ij}^{\text{\tiny eff}}$ (eq. \ref{eq:"CoupledParameters"}) within the slave rotor formalism reduces to a tight-binding hopping parameter $t$ throughout the system. In this limit the Hamiltonian support a perfect interference pattern that leads to localization. To give an example, a localized state $\psi_{\text{LS}}$ satisfies
\begin{equation}\label{eq:"Hop"}
    H^{\text{eff}}_i \psi_{\text{LS}}= t(d^\dagger_i d_{i+1} + d^\dagger_i d_{i+2}) \psi_{\text{LS}} = 0
\end{equation}
where $H^{\text{eff}}_i$ includes only the terms relevant to the destructive interference responsible for localization. However, as we turn on the interactions, the modifications within the Hamiltonian start to distort the interference pattern. The spatial variation within $h_i$ and $Z_i$ induces local fluctuations in the effective hopping strengths $t_{ij}^{\text{\tiny eff}}$, thereby perturbing the idealized interference condition. In this case, a strictly localized state with $\pm 1$ wavefunction does not necessarily satisfy the relation given in eq. \ref{eq:"Hop"}. Moreover, the local Lagrange multiplier shifts the energy of the LS and the spatial variance of it broadens its spectral feature as shown in Fig. \ref{Fig:"DOS_all"}, potentially enabling hybridization with bulk states. These mechanisms alter the structure of the LS, possibly leading to delocalization. In contrast to quasiperiodic systems, perfect destructive interference leading to Aharonov-Bohm caging can occur in periodic lattices such as the dice \cite{vidal1998aharonov,abilio1999magnetic} and rhombic lattices \cite{mukherjee2015observation}, where the interplay of geometry and magnetic flux leads to flat bands and localized states. However, interparticle interactions can disrupt this perfect interference, leading to band mixing and delocalization, as recently demonstrated in engineered flat-band Rydberg lattices \cite{chen2025interaction}.

We visualize in Fig. \ref{Fig:"LS_eg_A"} and \ref{Fig:"LS_eg_B"} the real-space spinon density \(\rho_i\) for four representative eigenstates. These states are selected to reflect both low and intermediate overlap with the non-interacting LS manifolds. Fig. \ref{Fig:"LS_eg_A"} (a, b) correspond to eigenstates with low and intermediate overlap \(\phi_A\) with the Type-A LS basis, while Fig. \ref{Fig:"LS_eg_B"} (a,b) show analogous states for the Type-B LS basis. As expected, the density profiles evolve from relatively delocalized and spatially extended structures at low \(\phi\), to more confined, LS-like patterns as \(\phi\) increases. However, even for intermediate overlap, the spatial support of these states often extends beyond the strictly localized support seen at \(U = 0\), indicating that while some LS character is preserved, the destructive interference that originally confined the state has been partially lifted. This spatial visualization complements the overlap metric and illustrates how correlations can gradually deform localized states in a nonuniform and site-dependent fashion.

\section{CONCLUSIONS}
We studied the Mott transition on AB quasicrystal using the slave-rotor mean-field theory. We constructed rational approximants of the AB tiling and identified two dominant types of localized states, Type-A and Type-B, arising from destructive interference in the non-interacting limit. By introducing Hubbard interactions, we mapped out the metal-Mott insulator phase diagram and analyzed the coordination-dependent modulation of the local quasiparticle weight \(Z_i\).

Focusing on the metallic regime, we studied the evolution of LS through the spinon spectrum and local observables. We developed an overlap metric to quantify the deformation of LS under interactions, revealing interaction-induced energy splitting, spectral broadening, and partial delocalization of the LS manifold. Type-B states, associated with higher coordination, consistently exhibited greater resilience to interaction effects.

Our results demonstrate that electron interactions in quasicrystals lead to rich, spatially inhomogeneous phenomena and selectively destabilize certain LS depending on their local geometric embedding. These findings highlight the importance of local connectivity in governing correlation effects in aperiodic systems and provide a foundation for further studies exploring flat-band phenomena and emergent correlated phases in quasiperiodic materials.

\section{ACKNOWLEDGMENTS}
This work is supported by TUBITAK 1001 program Grant no 122F346. The work at ASU was supported by the U.S. Department of Energy, Office of Science, Office of Basic
Energy Sciences, Material Sciences and Engineering Division under Award Number DE-SC0025247. E.Y. is supported by TUBITAK 2210 program.

\bibliography{ABL_slave_rotor_text}
\end{document}